\documentclass[twocolumn,showpacs]{revtex4}
\usepackage{graphicx,amssymb}
\usepackage{amsfonts}

\begin{document}
\title{A Complete Cosmological Solution\\
 to the Averaged Einstein Field Equations as found in Macroscopic Gravity\footnote{Copyright (2009) American Institute of Physics. This article may be downloaded for personal use only. Any other use requires prior permission of the author and the American Institute of Physics\\
 Article published in {\it J. Math. Phys.}, {\bf 50}, 082503 (2009) \\
 DOI: 10.1063/1.3193686\\
 URL: http://link.aip.org/link/?JMP/50/082503
 }}

\author{R. J. van den Hoogen}
\affiliation{Department of Mathematics, Statistics, and Computer Science\\
 St. Francis Xavier University\\
 Antigonish, Nova Scotia, Canada\\
 B2G 2W5}
\email{rvandenh@stfx.ca}
\date{July 10, 2009}
\begin{abstract}
A formalism for analyzing the complete set of field equations describing Macroscopic Gravity is presented.  Using this formalism, a cosmological solution to the Macroscopic Gravity equations is determined. It is found that if a particular segment of the connection correlation tensor is zero and if the macroscopic geometry is described by a flat Robertson-Walker metric, then the effective correction to the averaged Einstein Field equations of General Relativity i.e., the backreaction, is equivalent to a positive spatial curvature term.  This investigation completes the analysis of \cite{ColeyPelavasZalaletdinvov2005} and the formalism developed provides a possible basis for future studies.
\end{abstract}

\maketitle


\section{The Averaging Problem in Cosmology}

Modern cosmology is at an impasse. Observational studies indicate that our Universe on the largest of scales is highly isotropic (see \cite{WMAP} and references within), and appears to be expanding at an ever increasing rate (see \cite{DE} and references within).  Employing as our theory of gravity, Einstein's General Relativity, this scenario is modeled phenomenologically using a Friedman Robertson Walker(FRW) model with the addition of two new quantities and their consequent evolutions. There is no clear physical motivation for adding these two quantities.  Indeed, the existence of the two new quantities (``exotic'' forms of matter and/or energy), colloquially called Dark Matter and Dark Energy have yet to be observed, and in fact may never be directly observed.  However, the resulting cosmological model, called the $\Lambda$CDM model or Concordance model, fits the observed data remarkably well and is therefore widely accepted.  On the other hand, perhaps there is a simpler and more natural explanation. One that can help to explain the supposed effects of both Dark Matter and Dark Energy as the consequences of a single idea.  It is this idea that we explore here.

The most common assumption used in cosmology is that the universe is spatially homogeneous and isotropic on the largest of scales.  However, observational data shows that the universe is not spatially homogeneous nor isotropic at all scales since one observes structures.  In addition, these structures would not exist if the universe were spatially homogeneous and isotropic throughout its evolution.  Indeed, inhomogeneities in the early universe are required if we are to have any structure (stars, galaxies, clusters, filaments, walls, voids, etc.) at all at the present time.  With the assumption of spatial homogeneity and isotropy on the largest of scales, we have ignored any effects to the dynamical evolution of the universe that may be the direct result of inhomogeneities that we know are present on smaller scales, and at earlier times.  In fact, there has been some discussions in the literature argueing that the presently observed acceleration may be entirely due to the back-reaction of inhomogeneities found in our universe \cite{backreaction}.

In order to accurately reflect and model the effects of inhomogeneities over large space-time regions, one should employ an averaging procedure suitable for cosmological length scales. Indeed, Ellis \cite{Ellis} has argued that an averaged metric, and more importantly, an averaged set of field equations ought to be used when considering the evolution of large regions of space-time.  Given that General Relativity accurately models the local gravitational dynamics on much smaller scales, the development of an averaged theory should therefore be based on General Relativity.  The construction of an averaged theory of gravity based on Einstein's General Relativity is very difficult due to the non-linear
nature of the gravitational field equations. This problem is further exacerbated by the difficulty in defining a mathematically precise and covariant averaging procedure for tensor fields.  However, if these issues are resolved, then the resulting averaged theory of gravity will provide the theoretical framework needed to understand and model inhomogeneous cosmologies that are on average spatially homogeneous and isotropic on the largest of scales.

There have been many approaches to the averaging problem, (see \cite{Zalaletdinov,Zalaletdinov2,Zalaletdinov2008,Krasinski,OLD_AVERAGE,Futamase} and references within).
However, most of the approaches are either non-covariant, or perturbative in nature. One of the few approaches to the averaging problem that is both non-perturbative,
and covariant is the theory of Macroscopic Gravity as proposed and developed by R. Zalaletdinov  \cite{Zalaletdinov,Zalaletdinov2,Zalaletdinov2008}.
This approach employs a covariant space-time averaging procedure for tensor fields. When this averaging procedure is applied to
the field equations of General Relativity, one obtains a generalization of the Einstein Field Equations of General Relativity that includes a
new tensor field with its own set of field equations.

There has been some progress made in understanding the Zalaletdinov averaging procedure and its impact on the dynamics of the macroscopic spacetime \cite{ColeyPelavasZalaletdinvov2005,COVAR_SPAT,Coley_Pelavas,vandenHoogen2008}.  In \cite{COVAR_SPAT}, the authors started with a microscopic geometry, averaged using the Zalaletdinov averaging procedure and then made some assumptions to obtain the spatial averaging limit.  Once again in \cite{Coley_Pelavas}, the authors started with an inhomogeneous spherically symmetric spacetime and averaged using the Zalaletdinov procedure to explicitly calculate the form of the averaged Field Equations. Here we shall take an alternative point of view, the macroscopic point of view, where no explicit assumptions on the microscopic geometry are made, only that on average over large scales it can be described by the macroscopic geometry.  We will make some reasonable mathematical and geometrically plausible assumptions on the macroscopic geometry and the various correlation tensors and then we will analyze and solve the complete set of macroscopic gravity equations.  A previously published solution \cite{ColeyPelavasZalaletdinvov2005} did not provide a systematic and detailed procedure for obtaining a solution nor did it attempt to resolve the constraints arising from the existence of the affine deformation tensor.  Therefore, a complete, self-consistent solution to the Macroscopic Gravity equations appropriate for cosmological interpretations is presented below.

Regarding notation, covariant differentiation with respect to connection $\Gamma^\alpha{}_{\beta\gamma}$ is denoted with $||$. Round parentheses surrounding a set of indices represents symmetrization of the indices while square brackets represents anti-symmetrization.  Any index that is underlined is not included in the symmetrization.
We also define the totally antisymmetric tensor density $\varepsilon_{\alpha\beta\gamma\delta}=e_{\alpha\beta\gamma\delta}\sqrt{-\det(g_{\alpha\beta})}$ where the totally antisymmetric symbol is fixed so that $e_{1234}=1$.  We also define $\varepsilon_{\alpha\beta\gamma}=\varepsilon_{\alpha\beta\gamma\delta}u^{\delta}$.  Greek indices run from 1 to 4, while Latin indices run from 1 to 3. Units are chosen so that $c=G=1$, so $\kappa = 8\pi$.


\section{The Field Equations of Macroscopic Gravity}

\subsection{Brief Review of Macroscopic Gravity}

We shall use the same sign convention and similar notation as found in \cite{Zalaletdinov2,Zalaletdinov3} for most quantities.  A bar over a tensor or geometrical object indicates the spacetime averaged value of that quantity. Here we very briefly outline the general idea used in developing Zalaletdinov's theory of Macroscopic Gravity.  Assume our microscopic spacetime is a manifold $M$ with metric $g_{\alpha\beta}$, a Levi-Cevita connection $\gamma{}^{\alpha}{}_{\beta\gamma}$, and Riemann curvature tensor
\begin{equation}
r^{\alpha}{}_{\beta\gamma\delta}=2\gamma^\alpha{}_{\beta[\delta,\gamma]}+2 \gamma^\alpha{}_{\epsilon[\gamma} \gamma^\epsilon{}_{{\underline\beta}\delta]}.
\label{cartan1}
\end{equation}
Macroscopic Gravity is based on the idea that the average of the Levi-Cevita connection on $M$ yields a Levi-Cevita connection for the averaged or smoothed manifold $\overline M$, i.e., $\overline{ \gamma}\,{}^{\alpha}{}_{\beta\gamma}=\Gamma{}^{\alpha}{}_{\beta\gamma}$.  Given $\Gamma{}^{\alpha}{}_{\beta\gamma}$, by Frobeneius's theorem one can always determine locally a metric $G_{\alpha\beta}$ for $\overline M$ compatible with $\Gamma{}^{\alpha}{}_{\beta\gamma}$. Note this does not mean that $G_{\alpha\beta}$ is necessarily equal to the average of the metric $g_{\alpha\beta}$ found on the microscopic manifold $M$, i.e, $G_{\alpha\beta}\not={\overline g}_{\alpha\beta}$. The corresponding Riemann curvature tensor associated with $\Gamma{}^{\alpha}{}_{\beta\gamma}$ is $M^{\alpha}{}_{\beta\gamma\delta}=2\Gamma^\alpha{}_{\beta[\delta,\gamma]}+2 \Gamma^\alpha{}_{\epsilon[\gamma} \Gamma^\epsilon{}_{{\underline\beta}\delta]}$ with a Ricci tensor $M_{\alpha\beta}=M^{\mu}{}_{\alpha\beta\mu}$.  In Macroscopic Gravity, it is also assumed that the average of the microscopic Riemann curvature tensor, ${\overline r}\,^{\alpha}{}_{\beta\gamma\delta}=R^{\alpha}{}_{\beta\gamma\delta}$ yields another curvature tensor (non-Riemannian in nature) compatible with a second (necessarily non-metric) connection $\Pi{}^{\alpha}{}_{\beta\gamma}$.  Essential to the theory of Macroscopic Gravity is the definition of a connection correlation tensor
\begin{eqnarray}
Z{}^{\alpha}{}_{\beta\mu}{}^{\gamma}{}_{\delta\nu}&\equiv&\overline {(\gamma{}^{\alpha}{}_{\beta[\mu}\gamma{}^{\gamma}{}_{{\underline\delta}\nu]})}
-\overline{\gamma}\,{}^{\alpha}{}_{\beta[\mu}\,\overline{\gamma}\,{}^{\gamma}{}_{{\underline\delta}\nu]}\nonumber \\
&=&
\langle {\mathcal F}{}^{\alpha}{}_{\beta[\mu}{\mathcal F}{}^{\gamma}{}_{\underline{\delta}\nu]}\rangle-\langle{\mathcal F}{}^{\alpha}{}_{\beta[\mu}\rangle\langle{\mathcal F}{}^{\gamma}{}_{\underline{\delta}\nu]}\rangle
\end{eqnarray}
where ${\mathcal F}{}^\alpha{}_{\beta\gamma}$ is a bilocal extension of $\gamma{}^{\alpha}{}_{\beta\gamma}$ (see \cite{Zalaletdinov} for details).  The connection correlation tensor has index symmetries
\begin{eqnarray}
\mbox{Antisymmetric in $\mu$ and $\nu$}\qquad\quad\nonumber \\
Z^{\alpha}{}_{\beta\mu}{}^{\gamma}{}_{\delta\nu}&=&-Z^{\alpha}{}_{\beta\nu}{}^{\gamma}{}_{\delta\mu}\label{36}\\
\mbox{Antisymmetric in pairs ${\alpha \atop {\ \beta}}$ and ${\gamma \atop {\ \delta}}$}\nonumber\\
Z^{\alpha}{}_{\beta\mu}{}^{\gamma}{}_{\delta\nu}&=& -Z^{\gamma}{}_{\delta\mu}{}^{\alpha}{}_{\beta\nu}\label{pairs}
\end{eqnarray}
through construction. It is shown in \cite{Zalaletdinov,Zalaletdinov2} that this connection correlation tensor is able to provide the splitting rules necessary to determine the corrections required to successfully average out the Einstein Field Equations.  Various traces of $Z^{\alpha}{}_{\beta\mu}{}^{\gamma}{}_{\delta\nu}$ are needed and are defined as follows
\begin{eqnarray}
Z^{\alpha}{}_{\beta\delta\nu}&=& 2Z^{\alpha}{}_{\beta\epsilon}{}^{\epsilon}{}_{\delta\nu}\\
Q^{\alpha}{}_{\beta\mu\nu}&=&-2 Z^{\epsilon}{}_{\beta\mu}{}^{\alpha}{}_{\epsilon\nu}\\
Q_{\beta\mu}&=& Q^{\delta}{}_{\beta\mu\delta}
\end{eqnarray}
We shall assume that all higher order connection correlations except $Z^\alpha{}_{\beta\mu}{}^{\gamma}{}_{\delta\nu}$ are zero, this can be done in a self-consistent way as outlined in \cite{Zalaletdinov,Zalaletdinov2}.

\subsection{Timelike Vectors and Projections}

The average motion of the matter distribution and radiation in the universe defines a preferred direction.  We shall therefore adopt the conventional view that there is a well defined velocity vector of the matter distribution in the universe at least locally and through the Copernican principle, this then implies that there is a velocity vector at each point. We can therefore represent this average velocity by defining the unit timelike 4-velocity vector field
$$u^\alpha=\frac{dx^\alpha}{d\tau}, \qquad\qquad u^{\alpha}u_{\alpha}=-1$$ where $\tau$ is the proper time measured along the fundamental world-lines of the averaged matter distribution.  With this timelike vector field, and a given macroscopic metric $G_{\alpha\beta}$, one can now define the unique projection tensor
$$H_{\alpha\beta}=G_{\alpha\beta}+u_{\alpha}u_{\beta}$$ which projects into the 3-dimensional rest-space of observers moving with 4-velocity $u^{\alpha}$.

From now  on, a dot over an object represents the derivative of the object in the direction of vector field $u^{\alpha}$ that is, $\dot f = f_{||\alpha}u^{\alpha}$.   The covariant derivative of $u^{\alpha}$ can now be decomposed \cite{Stephani2004,EllisvanElst} as
\begin{equation}
u_{\alpha||\beta}=-{\dot u}_{\alpha}u_{\beta}+\omega_{\alpha\beta}+\sigma_{\alpha\beta}+\frac{1}{3}\theta H_{\alpha\beta}.
\end{equation}
The kinematic quantities are defined as the expansion $\theta=u^{\alpha}{}_{||\alpha}$, the shear, $\sigma_{\alpha\beta}= u_{(\alpha||\beta)}+ {\dot u}_{(\alpha}u_{\beta)} -\frac{1}{3}\theta H_{\alpha\beta}$, the vorticity $\omega_{\alpha\beta}=u_{[\alpha||\beta]}+{\dot u}_{[\alpha}u_{\beta]}$, and the acceleration ${\dot u}^{\alpha}=u^{\alpha}{}_{||\beta}u^{\beta}$.

The connection correlation tensor $Z{}^{\alpha}{}_{\beta\mu}{}^{\gamma}{}_{\delta\nu}$ is a $(2-2)$ valued 2-form that can be decomposed in an analogous fashion to the standard electro-magnetic form $F_{\alpha\beta}$.
We calculate two different projections (Electric and Magnetic) of $Z{}^{\alpha}{}_{\beta\mu}{}^{\gamma}{}_{\delta\nu}$ \cite{Zalaletdinov3} as follows
\begin{equation}
ZE{}^\alpha{}_\beta{}^\gamma{}_{\delta}{}_{\epsilon}= Z^\alpha{}_{\beta
\epsilon}{}^\gamma{}_{\delta \phi}u^{\phi}
\end{equation}
and
\begin{equation}
ZH{}^\alpha{}_\beta{}^\gamma{}_{\delta}{}_{\epsilon}= \frac{1}{2}Z^\alpha{}_{\beta \mu}
{}^\gamma{}_{\delta \nu}\varepsilon^{\mu \nu}{}_{ \epsilon \phi}u{^\phi}.
\end{equation}
The tensors $ZE{}^\alpha{}_\beta{}^\gamma{}_{\delta}{}_{\epsilon}$ and $ZH{}^\alpha{}_\beta{}^\gamma{}_{\delta}{}_{\epsilon}$ completely determine and characterize the correlation tensor $Z{}^{\alpha}{}_{\beta\mu}{}^{\gamma}{}_{\delta\nu}$, in the following manner
\begin{equation}
Z^\alpha{}_{\beta\mu}{}^\gamma{}_{\delta\nu}=-2ZE^\alpha{}_{\beta}{}^{\gamma}{}_{\delta[\mu}u_{\nu]} + ZH^\alpha{}_{\beta}{}^{\gamma}{}_{\delta\pi}\varepsilon^{\pi\chi}{}_{\mu\nu}u_{\chi}.
\end{equation}
We also note that $ZE{}^\alpha{}_\beta{}^\gamma{}_{\delta}{}_{\epsilon}$ and $ZH{}^\alpha{}_\beta{}^\gamma{}_{\delta}{}_{\epsilon}$ are antisymmetric in pairs
\begin{equation}
ZE{}^\alpha{}_\beta{}^\gamma{}_{\delta}{}_{\epsilon}=-ZE{}^\gamma{}_{\delta}{}^\alpha{}_\beta{}{}_{\epsilon},\qquad ZH{}^\alpha{}_\beta{}^\gamma{}_{\delta}{}_{\epsilon}=-ZH{}^\gamma{}_{\delta}{}^\alpha{}_\beta{}{}_{\epsilon},
\end{equation}
 and satisfy
\begin{equation}
ZE{}^\alpha{}_\beta{}^\gamma{}_{\delta}{}_{\epsilon}u^\epsilon=0, \qquad ZH{}^\alpha{}_\beta{}^\gamma{}_{\delta}{}_{\epsilon}u^\epsilon=0. \label{ortho}
\end{equation}
We are now able to count the number of independent components in $ZE{}^\alpha{}_\beta{}^\gamma{}_{\delta}{}_{\epsilon}$ and $ZH{}^\alpha{}_\beta{}^\gamma{}_{\delta}{}_{\epsilon}$: due to antisymmetry of pairs, and the spatial nature of the last index, we find that there are $\frac{(16)\cdot(16-1)}{2} \cdot 3 = 360$ independent components in each tensor $ZE{}^\alpha{}_\beta{}^\gamma{}_{\delta}{}_{\epsilon}$ and $ZH{}^\alpha{}_\beta{}^\gamma{}_{\delta}{}_{\epsilon}$ and consequently $Z^\alpha{}_{\beta\mu}{}^\gamma{}_{\delta\nu}$ has $720$ independent components in general.

\subsection{The Field Equations for the Connection Correlation Tensor}

\subsubsection{Algebraic Cyclic Identities}

By construction, the connection correlation tensor must satisfy the algebraic cyclic identity
\begin{equation}
Z^\alpha{}_{\beta[\mu}{}^{\gamma}{}_{\delta\nu]}=0 \Longleftrightarrow Z^\alpha{}_{\beta\mu}{}^{\gamma}{}_{\delta\nu}\varepsilon^{\mu\delta\nu\kappa}=0.\label{algebraic}
\end{equation}
Taking advantage of the decomposition of $Z{}^{\alpha}{}_{\beta\mu}{}^{\gamma}{}_{\delta\nu}$, one is able to rewrite the algebraic cyclic identities
as equivalent systems involving two equations through contractions with $u_{\kappa}$ to derive the first equation and with $H_{\kappa\chi}$ to obtain the second equation in each case.
\begin{eqnarray}
ZH^\alpha{}_{\beta}{}^{\gamma}{}_{\delta\mu}H^{\delta\mu}&=&0, \label{cyclic_2A} \\
ZH^\alpha{}_{\beta}{}^{\gamma}{}_{\delta\chi}u^{\delta}+
ZE^\alpha{}_{\beta}{}^{\gamma}{}_{\delta\mu}\varepsilon^{\delta\mu}{}_{\chi} &=& 0. \label{cyclic_2B}
\end{eqnarray}
We note that one can replace $H^{\delta\mu}$ with $G^{\delta\mu}$ in the above equations due to equation (\ref{ortho}). Equation (\ref{cyclic_2A}) yields $4^3=64$ constraints.
Equation (\ref{cyclic_2B}) can be further decomposed into a ``trace-like'' and a ``trace-free-like'' part.  Contracting equation (\ref{cyclic_2B}) with $H^{\beta\chi}$ and subtracting the result, one obtains the ``trace-free-like'' part (this is not totally trace-free in the sense that there are non-trivial traces in other indices),
\begin{eqnarray}
0&=& ZH^\alpha{}_{\beta}{}^{\gamma}{}_{\delta\chi}u^{\delta}+ ZE^\alpha{}_{\beta}{}^{\gamma}{}_{\delta\mu}\varepsilon^{\delta\mu}{}_{\chi}\nonumber\\
&&-\frac{1}{3}H_{\beta\chi}\left(ZH^\alpha{}_{\pi}{}^{\gamma}{}_{\delta\lambda}u^{\delta}H^{\pi\lambda} +ZE^\alpha{}_{\pi}{}^{\gamma}{}_{\delta\mu}\varepsilon^{\delta\mu\pi}\right) \label{cyclic_2D}
\end{eqnarray}
Using equations (\ref{ortho}) and (\ref{cyclic_2A}), the ``trace-like'' part reduces to a simple expression
\begin{equation}
ZE^\alpha{}_{\beta}{}^{\gamma}{}_{\delta\mu}\varepsilon^{\delta\mu\beta}=0. \label{cyclic_2C}
\end{equation}
The ``trace-free'' part, equation (\ref{cyclic_2D}), yields $(4^3)\cdot(3) - 16=176$ constraints, while the ``trace'' part, equation (\ref{cyclic_2C}), appears to yield $16$ constraints.  However, the ``trace'' part is actually symmetric in the remaining two indices due to antisymmetry in pairs, and therefore only yields $10$ linearly independent constraints.  Therefore, the total number of linearly independent constraints arising from the algebraic cyclic identity, equation (\ref{algebraic}), is $64+176+10=250$.

\subsubsection{Equi-affinity Constraint}

By assumption, the connection correlation tensor, $Z^\alpha{}_{\beta\mu}{}^{\gamma}{}_{\delta\nu}$ is assumed to satisfy
\begin{equation}
Z^\epsilon{}_{\epsilon\mu}{}^{\gamma}{}_{\delta\nu}=0.\label{equiaffine}
\end{equation}
In terms of $ZE{}^\alpha{}_\beta{}^\gamma{}_{\delta}{}_{\epsilon}$ and $ZH{}^\alpha{}_\beta{}^\gamma{}_{\delta}{}_{\epsilon}$ we have the following constraints
\begin{eqnarray}
ZH^\epsilon{}_{\epsilon}{}^{\gamma}{}_{\delta\mu} &=&0 \label{equiaffine_1}, \\
ZE^\epsilon{}_{\epsilon}{}^{\gamma}{}_{\delta\mu} &=& 0 \label{equiaffine_2}.
\end{eqnarray}
The number of constraints arising out of equations (\ref{equiaffine_1}) and (\ref{equiaffine_2}) is 45 each, due to the antisymmetry in pairs.  However, there are also $16$ of these constraints that are trivially satisfied due to equation (\ref{algebraic}).  Therefore the number of new linearly independent constraints arising from equation (\ref{equiaffine}) is $90-16=74$.

After all algebraic constraints are resolved, we find that there are $720-250-74=396$ independent components of the connection correlation tensor $Z^\alpha{}_{\beta\mu}{}^{\gamma}{}_{\delta\nu}$.

\subsubsection{Differential Cyclic Constraint}

By assumption, consistent with the assumption that all higher order connection correlations are zero, the connection correlation tensor is assumed to satisfy
\begin{equation}
Z^\alpha{}_{\beta[\mu}{}^{\gamma}{}_{{\underline\delta}\nu||\phi]}=0 \Longleftrightarrow
Z^\alpha{}_{\beta\mu}{}^{\gamma}{}_{\delta\nu||\phi}\varepsilon^{\mu\nu\phi\kappa}=0.\label{differential}
\end{equation}
This assumption is an integral component in the determination of a splitting rule for the average of the product of the Riemann tensor and the connection (see \cite{Zalaletdinov,Zalaletdinov2} for details).  Taking advantage of the decomposition of $Z^\alpha{}_{\beta\mu}{}^{\gamma}{}_{\delta\nu}$ one is able to rewrite the differential cyclic constraints as equivalent systems involving two equations through contractions with $u_{\kappa}$ to derive the first equation and with $H_{\kappa\chi}$ to obtain the second equation in each case.
\begin{eqnarray}
0 &=& ZH^\alpha{}_{\beta}{}^{\gamma}{}_{\delta\pi||\phi}H^{\pi\phi} +
2ZE^\alpha{}_{\beta}{}^{\gamma}{}_{\delta\mu} \omega^{\mu},\label{DE1}\\
0&=& \dot {ZH}{}^\alpha{}_{\beta}{}^{\gamma}{}_{\delta\kappa}H^{\kappa}{}_{\chi}
+ZH^\alpha{}_{\beta}{}^{\gamma}{}_{\delta\kappa}\left(\frac{2}{3}\theta H^{\kappa}{}_{\chi}-\sigma^{\kappa}{}_{\chi}-\omega^{\kappa}{}_{\chi}\right)\nonumber\\
&& \qquad -{ZE}^\alpha{}_{\beta}{}^{\gamma}{}_{\delta\mu}\dot u_\nu \varepsilon^{\mu\nu}{}_{\chi}
-ZE^\alpha{}_{\beta}{}^{\gamma}{}_{\delta\mu||\phi}\varepsilon^{\mu\phi}{}_{\chi} \label{DE2}
\end{eqnarray}
Note there is a dot over the first term in the second equation. Equations (\ref{DE1}) and (\ref{DE2}) are analogous to one half of the $1+3$ decomposition of the Maxwell Field Equations for $\bf E$ and $\bf H$ \cite{EllisvanElst}.

\subsubsection{The Quadratic Constraint}
The usual assumption in Macroscopic Gravity is to assume that the higher order connection correlations are identically zero.  If this is indeed the case then the connection correlation tensor $Z^\alpha{}_{\beta\mu}{}^{\gamma}{}_{\delta\nu}$ must also satisfy the quadratic constraint \cite{Zalaletdinov2}
\begin{eqnarray}
&&Z^{\delta }{}_{\beta \lbrack \gamma }{}^{\theta
}{}_{\underline{\kappa }\pi }Z^{\alpha }{}_{\underline{\delta
}\epsilon }{}^{\mu }{}_{\underline{\nu } \sigma ]}+
Z^{\delta
}{}_{\beta \lbrack \gamma }{}^{\mu }{}_{\underline{\nu } \sigma
}Z^{\theta }{}_{\underline{\kappa }\pi }{}^{\alpha }{}_{\underline{
\delta }\epsilon ]}+\nonumber 
\\
&&Z^{\alpha }{}_{\beta \lbrack \gamma }{}^{\delta }{}_{ \underline{\nu
}\sigma }Z^{\mu }{}_{\underline{\delta }\epsilon }{}^{\theta
}{}_{\underline{\kappa }\pi ]}+
Z^{\alpha }{}_{\beta \lbrack \gamma }{}^{\mu
}{}_{\underline{\delta } \epsilon }Z^{\theta }{}_{\underline{\kappa
}\pi }{}^{\delta }{}_{\underline{ \nu }\sigma ]}+\nonumber 
\\
&&Z^{\alpha
}{}_{\beta \lbrack \gamma }{}^{\theta }{}_{ \underline{\delta
}\epsilon }Z^{\mu }{}_{\underline{\nu }\sigma }{}^{\delta
}{}_{\underline{\kappa }\pi ]}+
Z^{\alpha }{}_{\beta \lbrack \gamma }{}^{\delta
}{}_{\underline{\kappa }\pi }Z^{\theta }{}_{\underline{\delta }
\epsilon }{}^{\mu }{}_{\underline{\nu }\sigma ]}=0.
\label{ZZ}
\end{eqnarray}

\subsection{The Field Equations for the Affine Deformation Tensor}

Taking the average of equation (\ref{cartan1}), and defining ${\overline r\,}^{\alpha}{}_{\beta\mu\nu} = R^{\alpha}{}_{\beta\mu\nu}$ to be another curvature tensor, one obtains a relationship between the Riemann curvature tensor, $M^\alpha{}_{\beta\mu\nu}$, the trace of the connection correlation tensor, $Q^\alpha{}_{\beta\mu\nu}$, and the non-Riemann curvature tensor $R^{\alpha}{}_{\beta\mu\nu}$ via
\begin{equation}
R^{\alpha}{}_{\beta\mu\nu}=M^{\alpha}{}_{\beta\mu\nu} + Q^{\alpha}{}_{\beta\mu\nu}.
\end{equation}
Given that $M^{\alpha}{}_{\beta[\mu\nu||\lambda]}=0$ (satisfies the Bianchi identity) and by assumption (\ref{differential}), we derive that $R^{\alpha}{}_{\beta[\mu\nu||\lambda]}=0$.  As stated earlier, there exists a second connection $\Pi^\alpha{}_{\beta\gamma}$ on the spacetime that generates the curvature tensor $R^{\alpha}{}_{\beta\rho\sigma}$.  However, it is considered easier to determine the difference between the metric connection $\Gamma^\alpha{}_{\beta\gamma}$ and the non-metric connection $\Pi^\alpha{}_{\beta\gamma}$.  We define the affine deformation tensor to be $A^\alpha{}_{\beta\gamma}=\Gamma^\alpha{}_{\beta\gamma}-\Pi^\alpha{}_{\beta\gamma}$ which must satisfy the following algebraic
equation \cite{Zalaletdinov,Zalaletdinov2}
\begin{equation}\label{AR_eq}
A^{\epsilon}{}_{\beta[\rho}R^{\alpha}{}_{\underline{\epsilon} \sigma
\lambda]}-A^{\alpha}{}_{\epsilon[\rho}R^{\epsilon}{}_{\underline{\beta}
\sigma \lambda]}=0,
\end{equation}
and the differential equation
\begin{equation}
A^{\alpha}{}_{\beta[\sigma||\rho]}-A^{\alpha}{}_{\epsilon[\rho}A^{\epsilon}{}_{\underline{\beta}\sigma]}=-\frac{1}{2}Q^{\alpha}{}_{\beta\rho\sigma}.\label{DA_eq}
\end{equation}

\subsection{The Averaged Einstein Field Equations}

The averaged Einstein field equations are
\begin{equation}
{\overline g\,}^{\alpha \epsilon }M_{\epsilon \beta }-\frac{1}{2}\delta _{\beta
}^{\alpha }{\overline g\,}^{\mu \nu }M_{\mu \nu }=-\kappa \langle {\bf t}_{\beta
}^{\alpha {\rm (micro)}}\rangle -\kappa T_{\beta }^{\alpha {\rm
(grav)}} \label{MG_eqns}
\end{equation}
where the averaged microscopic stress-energy tensor $\langle {\bf
t}_{\beta }^{\alpha {\rm (micro)}}\rangle $ is assumed to be described as a hydrodynamical fluid.  Once the components of $\bar{g}^{\mu \nu }$ and $Z^\alpha{}_{\beta\mu}{}^{\gamma}{}_{\delta\nu}$, are determined the gravitational stress-energy tensor $T_{\beta }^{\alpha {\rm (grav)}}$ due to connection correlations
\begin{equation}
(Z^{\alpha }{}_{\mu \nu \beta }-\frac{1}{2}\delta _{\beta }^{\alpha
}Q_{\mu \nu })\bar{g}^{\mu \nu }=-\kappa T_{\beta }^{\alpha {\rm
(grav)}}, \label{T_grav}
\end{equation}
can be calculated.  The effective energy density and isotropic pressure \cite{EllisvanElst} due to the connection correlation tensor are determined via
\begin{equation}
\rho_{grav}=T_{\beta }^{\alpha {\rm(grav)}}u_{\alpha}u{^\beta},\qquad
p_{grav}=\frac{1}{3}T_{\beta }^{\alpha {\rm(grav)}}H_{\alpha}{^\beta}.
\end{equation}
Rewriting we find
\begin{eqnarray}
\kappa\rho_{grav}&=&ZH^{\epsilon}{}_{\mu}{}^{\delta}{}_{\epsilon\pi}\varepsilon^{\pi\mu}{}_{\delta} \nonumber \\
&& \qquad +(2ZE^{\epsilon}{}_{\mu}{}^{\delta}{}_{\nu\epsilon}  +ZE^{\delta}{}_{\epsilon}{}^{\epsilon}{}_{\mu\nu})u_{\delta}G^{\mu\nu}\nonumber\\
&& \qquad\qquad +ZE^{\epsilon}{}_{\mu}{}^{\delta}{}_{\epsilon\delta}u^{\mu}\\
\kappa p_{grav}&=&\frac{2}{3}ZH^{\delta}{}_{\mu}{}^{\epsilon}{}_{\nu\pi}\varepsilon^{\pi}{}_{\delta\epsilon}G^{\mu\nu}-ZH^{\epsilon}{}_{\mu}{}^{\delta}{}_{\epsilon\pi}\varepsilon^{\pi\mu}{}_{\delta}\nonumber\\
&&\qquad -(\frac{2}{3}ZE^{\epsilon}{}_{\mu}{}^{\delta}{}_{\nu\epsilon} + ZE^{\delta}{}_{\epsilon}{}^{\epsilon}{}_{\mu\nu})u_{\delta}G^{\mu\nu}\nonumber\\&&\qquad\qquad-ZE^{\epsilon}{}_{\mu}{}^{\delta}{}_{\epsilon\delta}u^{\mu}
\end{eqnarray}
The momentum density and the anisotropic pressure due to the connection correlation tensor can be found in the usual manner as described in \cite{EllisvanElst}.


\section{Macroscopic Gravity in a Flat Spatially Homogeneous and Isotropic Spacetime}

\subsection{Assumptions} \label{assumptions}

Due to the difficulty in dealing with the extreme size of the tensorial objects involved in Macroscopic Gravity, we are enticed to make reasonable geometric assumptions about all objects involved in order to make some progress into the understanding of the structure of the Macroscopic Gravity equations. One must recall that only when geometric assumptions were made about the spacetime (i.e., spherically symmetric and static) in General Relativity did one make progress in obtaining solutions to the
Einstein Field Equations.

\paragraph*{\bf Assumption 1: The Averaged Metric and the Metric Correlations} We shall assume ${\overline g\,}_{\alpha\beta}=G_{\alpha\beta}$ and the following splitting rule $\overline{(g_{\alpha\beta}\gamma^{\gamma}{}_{\delta\epsilon})}=\overline{g}_{\alpha\beta}\overline{\gamma}^{\gamma}{}_{\delta\epsilon}$ which ensures that the spacetime average of the Christoffel symbol of the first kind for $M$ is the Christoffel symbol of the first kind for the averaged manifold $\overline M$. In addition, the metric correlations as outlined in \cite{Zalaletdinov,Zalaletdinov2} are also assumed to be zero. This way the average of the inverse microscopic metric is equal to the inverse of macroscopic metric, i.e, $\overline{(g_{\alpha\beta})^{-1}}\equiv\overline{g}^{\alpha\beta}=G^{\alpha\beta}$.  This should not be considered unusual in any spacetime with a very high degree of symmetry.

\paragraph*{\bf Assumption 2: Macroscopic Geometry}  For the present moment we are interested in investigating the equations of Macroscopic Gravity in $k=0$ Robertson Walker spacetimes.  The macroscopic metric, $G_{\alpha\beta}$, written in conformal coordinates has a line element of the form
\begin{equation}
ds^2 = G_{\alpha\beta}dx^\alpha dx^\beta = R^2(\eta)[-d\eta^2+dx^2+dy^2+dz^2].\label{metric}
\end{equation}
Conformal coordinates are used because they facilitate the calculations.  It is a straightforward exercise to transform to the usual cosmological coordinates with line element $ds^2 =-dt^2+R(t)^2(dx^2+dy^2+dz^2)$.  This spacetime is one of the simplest examples containing non-trivial curvature, connection, etc.  It is well known that the corresponding metric, equation (\ref{metric}), is invariant under the six-dimensional group (${\mathcal G}_{6}$) of Killing vectors, generated by $\{{\bf P}_i,{\bf M}_{ij}\}$, spatial translations and spatial rotations, \cite{MaartensMaharaj86} where
\begin{eqnarray*}
{\bf P}_i &=& \partial_i \\
{\bf M}_{ij}&=&x_i\partial_j -x_j\partial_i
\end{eqnarray*}
This spacetime also admits a timelike vector field orthogonal to the spatial hyper-surfaces of homogeneity and isotropy,
$u^\alpha = [-\frac{1}{R(\eta)},0,0,0]$. This unit timelike vector has zero acceleration, zero vorticity and zero shear, and therefore $u_{\alpha||\beta}=\frac{1}{3}\theta H_{\alpha\beta}$ where $\theta=3\frac{R'(\eta)}{R(\eta)^2}$.

\paragraph*{\bf Assumption 3: Invariance of Macroscopic Gravity Objects}    We shall also assume that the connection correlation tensor $Z^\alpha{}_{\beta\mu}{}^{\gamma}{}_{\delta\nu}$ and the affine deformation tensor $A^{\alpha}{}_{\beta\gamma}$ are invariant under the same ${\mathcal G}_{6}$ of Killing vectors as the macroscopic metric.  These assumptions ensure that $Z^\alpha{}_{\beta\mu}{}^{\gamma}{}_{\delta\nu}$ and $A^{\alpha}{}_{\beta\gamma}$ are compatible with the geometry of the macroscopic spacetime.  Note, $Z^\alpha{}_{\beta\mu}{}^{\gamma}{}_{\delta\nu}$ and $A^{\alpha}{}_{\beta\gamma}$ {\bf need not} be invariant under the same six dimensional group of Killing vectors, ${\mathcal G}_6$,  as the macroscopic spacetime apriori.  The constraints are then
\begin{eqnarray}
{\mathcal L}_{\bf X} {\bf G} &=& 0,    \label{Lie_g}\\
{\mathcal L}_{\bf X} {\bf Z} &=& 0,    \label{Lie_Z}\\
{\mathcal L}_{\bf X} {\bf A} &=& 0,    \label{Lie_A}
\end{eqnarray}
 where ${\bf X} \in  {\mathcal G}_{6}$. Equation (\ref{Lie_g}) is just assumption {\bf 2} while equations (\ref{Lie_Z}) and (\ref{Lie_A}) will place significant
constraints on the number of independent components of $Z^\alpha{}_{\beta\mu}{}^{\gamma}{}_{\delta\nu}$ and $A^{\alpha}{}_{\beta\gamma}$.  We shall also see later, that equations (\ref{Lie_Z}) and (\ref{Lie_A}) can be relaxed and some conclusions can still be made.

\paragraph*{\bf Assumption 4: Averaged Microscopic Matter}  We shall also assume that the averaged microscopic energy momentum tensor can be modeled macroscopically as a perfect fluid, that is
\begin{equation}
\langle {\bf t}^{\alpha {\rm \ (micro)}}_\beta\rangle = \rho_{mat}u^\alpha u_\beta+p_{mat}H^\alpha{}_{\beta}
\end{equation} where $\rho_{mat}$ and $p_{mat}$ are the energy density and pressure for the averaged matter and where $u^{\alpha}$ can now also be interpreted as the average four-velocity of the fluid.

\paragraph*{\bf Assumption 5: Electric part of the Connection Correlation Tensor}  We shall assume that the electric part of the connection correlation tensor is zero, that is, $ZE^\alpha{}_{\beta}{}^{\gamma}{}_{\delta\mu}=0$. The primary motivation for setting $ZE^\alpha{}_{\beta}{}^{\gamma}{}_{\delta\mu}=0$ is that the constraint equation (\ref{ZZ}) is identically satisfied.

One should note that placing constraints on the connection correlation tensor such as assumptions {\bf 3} and {\bf 5} is equivalent to prescribing some rule to split the average of the product of connections and the product of the averages of the connection for the microscopic geometry (i.e., splitting rules for products of the connection).  For example, as an illustration only, if one was to assume that the microscopic geometry was described by an almost FRW metric of the form $ds^2=R(\eta)^2[-(1+2\Phi)d\eta^2+(1-2\Psi)(dx^2+dy^2+dz^2)]$ where $\Phi(\eta,x,y,z)$ and $\Psi(\eta,x,y,z)$, then assumption {\bf 5} imposes many relations involving splitting rules, including for example
 \begin{widetext}
 $$ ZE^0{}_{1}{}^{1}{}_{12}=Z^{0}{}_{12}{}^{1}{}_{10}=\overline{\gamma^{0}_{1[2}\gamma^{1}_{\underline{1}0]}}-\overline{\gamma^{0}_{1[2}}\,\,\overline{\gamma^{1}_{\underline{1}0]}}=0
 \Rightarrow \overline{\frac{\Phi_x \Psi_y}{(1+2\Phi)(1-2\Psi)}}=\overline{\frac{\Phi_x}{1+2\Phi}}\,\,\overline{\frac{\Psi_y}{1-2\Psi}}.$$
 \end{widetext}
However, not all of the products of connections are split trivially nor are they fixed by the assumptions above, for example,
\begin{widetext}
$$ZB^{1}{}_{2}{}^{1}{}_{31}=Z^{1}{}_{22}{}^{1}{}_{33}=\overline{\gamma^{1}{}_{2[2}\gamma^{1}{}_{\underline{3}3]}} -\overline{\gamma^{1}{}_{2[2}}\,\,\overline{\gamma^{1}{}_{\underline{3}3]}} \not= 0
\Rightarrow \overline{\left(\frac{\Psi_x}{1-2\Psi}\right)^2}\not=\left(\overline{\frac{\Psi_x}{1-2\Psi}}\right)^2.$$
\end{widetext}
This paragraph has only been added to illustrate the effect these assumptions have on the connection correlation tensor given a simple inhomogeneous microscopic geometry.  From now on, we no longer assume anything about the microscopic geometry, only that it averages on sufficiently large scales to the macroscopic geometry described by assumption {\bf 2}.

The assumptions {\bf 1}, thru {\bf 5} greatly simplify the situation, allowing one to completely solve the Macroscopic Gravity equations as presented in \cite{Zalaletdinov,Zalaletdinov2}.  The software package GRTensor II \cite{GRTENSOR} is used to do and check all calculations that follow.

\subsection{The Connection Correlation Tensor}

\subsubsection{Equivalence Classes}

We shall now assume that $Z^{\alpha}{}_{\beta\mu}{}^{\gamma}{}_{\delta\nu}$ satisfies equations (\ref{36}) and (\ref{pairs}) and is invariant under the same Lie group of motions as the macroscopic metric.  $ZE{}^\alpha{}_\beta{}^\gamma{}_{\delta}{}_{\epsilon}$ and $ZH{}^\alpha{}_\beta{}^\gamma{}_{\delta}{}_{\epsilon}$ are then also invariant under the same group of motions.  Each component of $ZH{}^\alpha{}_\beta{}^\gamma{}_{\delta}{}_{\epsilon}$ and $ZE{}^\alpha{}_\beta{}^\gamma{}_{\delta}{}_{\epsilon}$ is a function of $\eta$ only. The components of $Z^{\alpha}{}_{\beta\mu}{}^{\gamma}{}_{\delta\nu}$ can be classified into 40 equivalence classes; 20 classes of components in each of $ZE{}^\alpha{}_\beta{}^\gamma{}_{\delta}{}_{\epsilon}$ and $ZH{}^\alpha{}_\beta{}^\gamma{}_{\delta}{}_{\epsilon}$, not all of which are linearly independent.  In what follows $[ijk]$ is one element of the ordered triple $\{123,312,231\}$, and $[jk]$ or $[ij]$ is one element of the order pair $\{12,13,23\}$. In any class that contains a single item, the number of nontrivial components of  $Z^{\alpha}{}_{\beta\mu}{}^{\gamma}{}_{\delta\nu}$ in that class is 12, while in classes containing two items, the number of components is 24, 12 for each element. The notation used here and that used in \cite{Zalaletdinov3} are slightly different.

We define the $\mathcal H$ and $\mathcal B$ - equivalence classes as the following 20 sets of equivalent components of $ZH{}^\alpha{}_\beta{}^\gamma{}_{\delta}{}_{\epsilon}$.
$$\begin{tabular}{ll} 
${\mathcal H}_1 = \{ZH^{j}{}_{j}{}^{k}{}_{ji}, ZH^{j}{}_{k}{}^{k}{}_{ki}\}$    & ${\mathcal B}_1 = \{ZH^{j}{}_{i}{}^{j}{}_{0i},ZH^{i}{}_{j}{}^{i}{}_{0j} \}$\\
${\mathcal H}_2 = \{ZH^{j}{}_{j}{}^{j}{}_{ki}, ZH^{k}{}_{j}{}^{k}{}_{ki}\}$    & ${\mathcal B}_2 = \{ZH^{j}{}_{j}{}^{i}{}_{0i},ZH^{i}{}_{i}{}^{j}{}_{0j} \}$\\
${\mathcal H}_3 = \{ZH^{i}{}_{j}{}^{k}{}_{ii}, ZH^{j}{}_{i}{}^{i}{}_{ki}\}$    & ${\mathcal B}_3 = \{ZH^{i}{}_{j}{}^{j}{}_{0i},ZH^{j}{}_{i}{}^{i}{}_{0j} \}$\\
${\mathcal H}_4 = \{ZH^{i}{}_{i}{}^{j}{}_{ki}, ZH^{k}{}_{j}{}^{i}{}_{ii}\}$    & ${\mathcal B}_4 = \{ZH^{0}{}_{0}{}^{i}{}_{0i} \}$\\
${\mathcal H}_5 = \{ZH^{0}{}_{j}{}^{k}{}_{0i}, ZH^{j}{}_{0}{}^{0}{}_{ki}\}$    & ${\mathcal B}_5 = \{ZH^{j}{}_{i}{}^{0}{}_{ji},ZH^{i}{}_{j}{}^{0}{}_{ij} \}$\\
${\mathcal H}_6 = \{ZH^{0}{}_{0}{}^{j}{}_{ki}, ZH^{k}{}_{j}{}^{0}{}_{0i}\}$    & ${\mathcal B}_6 = \{ZH^{j}{}_{j}{}^{0}{}_{ii},ZH^{i}{}_{i}{}^{0}{}_{jj} \}$\\
${\mathcal H}_7 = \{ZH^{j}{}_{0}{}^{k}{}_{0i} \}$                              & ${\mathcal B}_7 = \{ZH^{i}{}_{j}{}^{0}{}_{ji},ZH^{j}{}_{i}{}^{0}{}_{ij}  \}$\\
${\mathcal H}_8 = \{ZH^{0}{}_{j}{}^{0}{}_{ki} \}$                              & ${\mathcal B}_8 = \{ZH^{0}{}_{0}{}^{0}{}_{ii} \}$\\
${\mathcal H}_9 = \{ZH^{i}{}_{j}{}^{i}{}_{ki} \}$                              & ${\mathcal B}_9 = \{ZH^{i}{}_{i}{}^{i}{}_{0i} \}$\\
${\mathcal H}_{10} = \{ZH^{j}{}_{i}{}^{k}{}_{ii}\}$                            & ${\mathcal B}_{10} = \{ZH^{i}{}_{i}{}^{0}{}_{ii} \}$
\end{tabular}$$

Although it is not necessary for the paper, for completeness, we define the $\mathcal E$ and $\mathcal D$ - equivalence classes as the following 20 sets of equivalent components of $ZE{}^\alpha{}_\beta{}^\gamma{}_{\delta}{}_{\epsilon}$.
$$\begin{tabular}{ll} 
${\mathcal E}_1 = \{ZE^{j}{}_{j}{}^{k}{}_{ji}, ZE^{j}{}_{k}{}^{k}{}_{ki}\}$      & ${\mathcal D}_1 = \{ZE^{j}{}_{i}{}^{j}{}_{0i},ZE^{i}{}_{j}{}^{i}{}_{0j} \}$\\
${\mathcal E}_2 = \{ZE^{j}{}_{j}{}^{j}{}_{ki}, ZE^{k}{}_{j}{}^{k}{}_{ki}\}$      & ${\mathcal D}_2 = \{ZE^{j}{}_{j}{}^{i}{}_{0i},ZE^{i}{}_{i}{}^{j}{}_{0j} \}$\\
${\mathcal E}_3 = \{ZE^{i}{}_{j}{}^{k}{}_{ii}, ZE^{j}{}_{i}{}^{i}{}_{ki}\}$      & ${\mathcal D}_3 = \{ZE^{i}{}_{j}{}^{j}{}_{0i},ZE^{j}{}_{i}{}^{i}{}_{0j} \}$\\
${\mathcal E}_4 = \{ZE^{i}{}_{i}{}^{j}{}_{ki}, ZE^{k}{}_{j}{}^{i}{}_{ii}\}$      & ${\mathcal D}_4 = \{ZE^{0}{}_{0}{}^{i}{}_{0i} \}$\\
${\mathcal E}_5 = \{ZE^{0}{}_{j}{}^{k}{}_{0i}, ZE^{j}{}_{0}{}^{0}{}_{ki}\}$      & ${\mathcal D}_5 = \{ZE^{j}{}_{i}{}^{0}{}_{ji},ZE^{i}{}_{j}{}^{0}{}_{ij} \}$\\
${\mathcal E}_6 = \{ZE^{0}{}_{0}{}^{j}{}_{ki}, ZE^{k}{}_{j}{}^{0}{}_{0i}\}$      & ${\mathcal D}_6 = \{ZE^{j}{}_{j}{}^{0}{}_{ii},ZE^{i}{}_{i}{}^{0}{}_{jj} \}$\\
${\mathcal E}_7 = \{ZE^{j}{}_{0}{}^{k}{}_{0i} \}$                                & ${\mathcal D}_7 = \{ZE^{i}{}_{j}{}^{0}{}_{ji},ZE^{j}{}_{i}{}^{0}{}_{ij}  \}$\\
${\mathcal E}_8 = \{ZE^{0}{}_{j}{}^{0}{}_{ki} \}$                                & ${\mathcal D}_8 = \{ZE^{0}{}_{0}{}^{0}{}_{ii} \}$\\
${\mathcal E}_9 = \{ZE^{i}{}_{j}{}^{i}{}_{ki} \}$                                & ${\mathcal D}_9 = \{ZE^{i}{}_{i}{}^{i}{}_{0i} \}$\\
${\mathcal E}_{10} = \{ZE^{j}{}_{i}{}^{k}{}_{ii}\}$                              & ${\mathcal D}_{10} = \{ZE^{i}{}_{i}{}^{0}{}_{ii} \}$
\end{tabular}$$

Note, that there are 40 equivalence classes of components of $Z^{\alpha}{}_{\beta\mu}{}^{\gamma}{}_{\delta\nu}$, but only 32 of them are independent.  There are additional relationships between some of the components within some of the different equivalence classes.  This complication can be managed by carefully choosing the independent variables.

\begin{widetext}
\subsubsection{Variables}

As described above there are only 32 independent components of $Z^{\alpha}{}_{\beta\mu}{}^{\gamma}{}_{\delta\nu}$; 16 independent components in each of $ZH{}^\alpha{}_\beta{}^\gamma{}_{\delta}{}_{\epsilon}$ and $ZE{}^\alpha{}_\beta{}^\gamma{}_{\delta}{}_{\epsilon}$.  Alternatively, there are 8 independent components in each of the four equivalence classes $\mathcal H,B,E,D$.  Here we make the choice as to what we will call our primary independent variables.  Note $R(\eta)^2$ is the metric function which is added to the definition of our variables strictly for convenience.

\begin{eqnarray*} 
H_1(\eta) R(\eta)^2= ZH^{j}{}_{j}{}^{k}{}_{ji} \in {\mathcal H}_1 \qquad B_1(\eta) R(\eta)^2= ZH^{j}{}_{i}{}^{j}{}_{0i} \in {\mathcal B}_1\\
H_2(\eta) R(\eta)^2= ZH^{j}{}_{j}{}^{j}{}_{ki} \in {\mathcal H}_2 \qquad B_2(\eta) R(\eta)^2= ZH^{j}{}_{j}{}^{i}{}_{0i} \in {\mathcal B}_2\\
H_3(\eta) R(\eta)^2= ZH^{i}{}_{j}{}^{k}{}_{ii} \in {\mathcal H}_3 \qquad B_3(\eta) R(\eta)^2= ZH^{i}{}_{j}{}^{j}{}_{0i} \in {\mathcal B}_3\\
H_4(\eta) R(\eta)^2= ZH^{i}{}_{i}{}^{j}{}_{ki} \in {\mathcal H}_4 \qquad B_4(\eta) R(\eta)^2= ZH^{0}{}_{0}{}^{i}{}_{0i} \in {\mathcal B}_4\\
H_5(\eta) R(\eta)^2= ZH^{0}{}_{j}{}^{k}{}_{0i} \in {\mathcal H}_5 \qquad B_5(\eta) R(\eta)^2= ZH^{j}{}_{i}{}^{0}{}_{ji} \in {\mathcal B}_5\\
H_6(\eta) R(\eta)^2= ZH^{0}{}_{0}{}^{j}{}_{ki} \in {\mathcal H}_6 \qquad B_6(\eta) R(\eta)^2= ZH^{j}{}_{j}{}^{0}{}_{ii} \in {\mathcal B}_6\\
H_7(\eta) R(\eta)^2= ZH^{j}{}_{0}{}^{k}{}_{0i} \in {\mathcal H}_7 \qquad B_7(\eta) R(\eta)^2= ZH^{i}{}_{j}{}^{0}{}_{ji} \in {\mathcal B}_7\\
H_8(\eta) R(\eta)^2= ZH^{0}{}_{j}{}^{0}{}_{ki} \in {\mathcal H}_8 \qquad B_8(\eta) R(\eta)^2= ZH^{0}{}_{0}{}^{0}{}_{ii} \in {\mathcal B}_8
\end{eqnarray*}

\begin{eqnarray*} 
E_1(\eta) R(\eta)^2= ZE^{j}{}_{j}{}^{k}{}_{ji} \in {\mathcal E}_1 \qquad D_1(\eta) R(\eta)^2= ZE^{j}{}_{i}{}^{j}{}_{0i} \in {\mathcal D}_1\\
E_2(\eta) R(\eta)^2= ZE^{j}{}_{j}{}^{j}{}_{ki} \in {\mathcal E}_2 \qquad D_2(\eta) R(\eta)^2= ZE^{j}{}_{j}{}^{i}{}_{0i} \in {\mathcal D}_2\\
E_3(\eta) R(\eta)^2= ZE^{i}{}_{j}{}^{k}{}_{ii} \in {\mathcal E}_3 \qquad D_3(\eta) R(\eta)^2= ZE^{i}{}_{j}{}^{j}{}_{0i} \in {\mathcal D}_3\\
E_4(\eta) R(\eta)^2= ZE^{i}{}_{i}{}^{j}{}_{ki} \in {\mathcal E}_4 \qquad D_4(\eta) R(\eta)^2= ZE^{0}{}_{0}{}^{i}{}_{0i} \in {\mathcal D}_4\\
E_5(\eta) R(\eta)^2= ZE^{0}{}_{j}{}^{k}{}_{0i} \in {\mathcal E}_5 \qquad D_5(\eta) R(\eta)^2= ZE^{j}{}_{i}{}^{0}{}_{ji} \in {\mathcal D}_5\\
E_6(\eta) R(\eta)^2= ZE^{0}{}_{0}{}^{j}{}_{ki} \in {\mathcal E}_6 \qquad D_6(\eta) R(\eta)^2= ZE^{j}{}_{j}{}^{0}{}_{ii} \in {\mathcal D}_6\\
E_7(\eta) R(\eta)^2= ZE^{j}{}_{0}{}^{k}{}_{0i} \in {\mathcal E}_7 \qquad D_7(\eta) R(\eta)^2= ZE^{i}{}_{j}{}^{0}{}_{ji} \in {\mathcal D}_7\\
E_8(\eta) R(\eta)^2= ZE^{0}{}_{j}{}^{0}{}_{ki} \in {\mathcal E}_8 \qquad D_8(\eta) R(\eta)^2= ZE^{0}{}_{0}{}^{0}{}_{ii} \in {\mathcal D}_8
\end{eqnarray*}

It should be noted for completeness and clarity the following eight relations detailing the relationships within the other eight equivalence classes.
\begin{eqnarray*}
\left(H_2(\eta)-H_3(\eta)-H_4(\eta)\right)R(\eta)^2 &=& ZH^{i}{}_{j}{}^{i}{}_{ki} \in {\mathcal H}_9  \\
\left(H_1(\eta)-H_3(\eta)+H_4(\eta)\right)R(\eta)^2 &=& ZH^{j}{}_{i}{}^{k}{}_{ii} \in {\mathcal H}_{10} \\
\left(B_1(\eta)+B_2(\eta)+B_3(\eta)\right)R(\eta)^2 &=& ZH^{i}{}_{i}{}^{i}{}_{0i} \in {\mathcal B}_9 \\
\left(B_5(\eta)+B_6(\eta)+B_7(\eta)\right)R(\eta)^2 &=& ZH^{i}{}_{i}{}^{0}{}_{ii} \in {\mathcal B}_{10} \\
\left(E_2(\eta)-E_3(\eta)-E_4(\eta)\right)R(\eta)^2 &=& ZE^{i}{}_{j}{}^{i}{}_{ki} \in {\mathcal E}_9  \\
\left(E_1(\eta)-E_3(\eta)+E_4(\eta)\right)R(\eta)^2 &=& ZE^{j}{}_{i}{}^{k}{}_{ii} \in {\mathcal E}_{10} \\
\left(D_1(\eta)+D_2(\eta)+D_3(\eta)\right)R(\eta)^2 &=& ZE^{i}{}_{i}{}^{i}{}_{0i} \in {\mathcal D}_9 \\
\left(D_5(\eta)+D_6(\eta)+D_7(\eta)\right)R(\eta)^2 &=& ZE^{i}{}_{i}{}^{0}{}_{ii} \in {\mathcal D}_{10} \\
\end{eqnarray*}

Now that we have carefully chosen our independent variables, we can proceed onto displaying and solving the equations of macroscopic gravity.  From now on, we shall assume that $ZE{}^\alpha{}_\beta{}^\gamma{}_{\delta}{}_{\epsilon}=0$ which is easily accomplished by setting $D_i(\eta)=E_i(\eta)=0$.  We shall no longer explicitly write out the functional dependencies in $H_i(\eta)$, $B_i(\eta)$ and $R(\eta)$ unless there is a risk of ambiguity.

\subsection{The Field Equations for the Connection Correlation Tensor}

Equation (\ref{cyclic_2A}) yields
\begin{eqnarray}
H_1-2H_3+2H_4 &=& 0 \nonumber\\
3B_1+B_2+B_3 &=& 0 \nonumber\\
3B_5+B_6+B_7 &=& 0\nonumber\\
B_5+3B_6+B_7 &=& 0\nonumber\\
B_8 &=& 0
\end{eqnarray}
Equation (\ref{cyclic_2D}) yields
\begin{eqnarray}
H_5 &=& 0\nonumber\\
H_6 &=& 0\nonumber\\
H_7 &=& 0\nonumber\\
B_2 &=& 0\nonumber\\
B_3  &=& 0\nonumber\\
B_2+B_3  &=& 0\nonumber\\
B_4 &=& 0
\end{eqnarray}
Equation (\ref{cyclic_2C}) is identically satisfied.
Equation (\ref{equiaffine}) yields
\begin{eqnarray}\label{Equi-affine}
-H_1+H_2+H_4+H_6 &=& 0 \nonumber\\
B_1+3B_2+B_3+B_4 &=& 0 \nonumber\\
B_5+3B_6+B_7+B_8 &=& 0
\end{eqnarray}

Equation (\ref{DE1}) yields
\begin{eqnarray}
\frac{R'}{R^2}\left(H_1-2H_3+2H_4+H_5-H_6+H_7\right) &=& 0 \\
\frac{R'}{R^2}\left(H_2-2H_3-2H_4+H_5+H_6+H_8\right) &=& 0 \\
\frac{R'}{R^2}\left(B_1+B_2+3B_3-B_4+3B_5+B_6+B_7-B_8\right) &=& 0 \\
\frac{R'}{R^2}\left(B_1+3B_2+B_3+B_4-B_5-3B_6-B_7-B_8\right) &=& 0
\end{eqnarray}

The resulting non-trivial differential equations from equation (\ref{DE2}) are
\begin{eqnarray}
H_i'  &=& -2\frac{R' }{R } H_i\nonumber\\
B_i'  &=& -2\frac{R' }{R } B_i \label{DE3}
\end{eqnarray}

The solution to the algebraic constraints can be parameterized as
\begin{equation}
\left[H_1,H_2,H_3,H_4,H_5,H_6,H_7,H_8\right] =
{\mathcal H}_1\left[2 , 2 , 1 , 0 , 0 , 0 , 0 , 0\right] +
{\mathcal H}_2\left[0 , 1 , -1 , -1 , 0 , 0 , 0, -5\right]
\label{sol1}\end{equation}
and
\begin{equation}
\left[B_1,B_2,B_3,B_4,B_5,B_6,B_7,B_8\right] =
{\mathcal B}_1\left[0 , 0 ,0 , 0, 1 , 1 , -4 ,0 \right]\label{sol2}
\end{equation}
where each of the parameters ${\mathcal H}_1,{\mathcal H}_2,$ and ${\mathcal B}_1$ are functions of $\eta$.  The number of independent variables is now reduced to three.  The only remaining equation to be solved are the differential cyclic constraints, equation (\ref{DE3}), which are easily integrated to yield
\begin{equation}{\mathcal H}_1(\eta)=\frac{h_1}{R^2(\eta)}, \quad {\mathcal H}_2(\eta)=\frac{h_2}{R^2(\eta)}, \quad {\mathcal B}_1(\eta)=\frac{b_1}{R^2(\eta)}\label{sol3}
\end{equation}
where $h_1,h_2$ and $b_1$ are arbitrary constants.  The field equations for the connection correlation tensor have been completely resolved to reveal that each non-trivial component of $Z^\alpha{}_{\beta\mu}{}^{\gamma}{}_{\delta\nu}$ is constant in this system of coordinates.  This is precisely the ansatz used in deriving the solution found in \cite{ColeyPelavasZalaletdinvov2005}.

\subsection{The Field Equations for the Affine Deformation Tensor}

Here, the affine deformation tensor $A^\alpha{}_{\beta\gamma}$ is also assumed to be invariant under the ${\mathcal G}_6$ group of motions.  This leaves three independent equivalence classes of components undetermined.  We define our variables to be the following
\begin{eqnarray*}
A_1(\eta)R(\eta) &=& A^0{}_{ii},\\
A_2(\eta)R(\eta)&=&A^i{}_{i0},\\
A_3(\eta)R(\eta)&=&A^0{}_{0 0}
\end{eqnarray*}
Equation (\ref{AR_eq}) yields
\begin{eqnarray}
0 &=& (B_1+B_2+3B_3-B_4)A_1+ (3B_5+B_6+B_7-B_8)A_2 ,\nonumber\\
0 &=&(3B_5+B_6+B_7-B_8)A_2 - (3B_5+B_6+B_7-B_8)A_3 ,\nonumber\\
0 &=& (B_1+B_2+3B_3-B_4)A_2 -(B_1+B_2+3B_3-B_4)A_3,\nonumber\\
0 &=&\left[\frac{R''}{R^3}-\frac{(R')^2}{R^4}\right](A_1+A_2) .\label{AR2a}
\end{eqnarray}
Equation(\ref{DA_eq}) yields
\begin{eqnarray}
0&=& B_1+B_2+3B_3-B_4,\nonumber\\
0&=& 3B_5+B_6+B_7-B_8,\nonumber\\
0&=& \frac{R'}{R^2}(A_1+A_2)-A_1A_2+2(H_1+H_2+H_3+H_5),\nonumber\\
\frac{A_1'}{R} &=& -\frac{R'}{R^2}A_1 + \frac{R'}{R^2}(A_2 -A_3) - A_1(A_2- A_3),\nonumber\\
\frac{A_2'}{R} &=& -\frac{R'}{R^2}A_2 - \frac{R'}{R^2}(A_2-A_3) + A_2(A_2 -A_3).
\end{eqnarray}
\end{widetext}

After substituting the solution given by equations (\ref{sol1},\ref{sol2},\ref{sol3}) we note that the equations containing the $B_i$'s are trivially satisfied, and the remaining term arising from the connection correlation tensor becomes $H_1+H_2+H_3+H_5=5h_1/R^2$.  Equation (\ref{AR2a}) has a solution $A_1 = -A_2$ unless $R=R_0e^{\alpha\eta}$ where $R_0$ and $\alpha$ are two arbitrary constants of integration (Milne solution).  If one was to pursue this possibility, and proceeds to solve the averaged Einstein Field Equations in the next section, one finds that $p_{mat}=-\frac{1}{3}\rho_{mat}\propto R^{-2}$, that is, the matter can not be prescribed and must behave as curvature, no other forms of matter (dust, radiation) are permitted.   Therefore, the only plausible solution is obtained by letting $A_1=-A_2$.  In this case we easily solve the remaining equations and find that
\begin{eqnarray*}
h_1 &=& -\frac{1}{10}{\mathcal A}^2,\nonumber\\
A_{1}(\eta) &=& \frac{\mathcal A}{R(\eta)}, \nonumber\\
A_{2}(\eta) &=& -\frac{\mathcal A}{R(\eta)}, \nonumber\\
A_{3}(\eta) &=& -\frac{\mathcal A}{R(\eta)},\label{sol4}
\end{eqnarray*}
where ${\mathcal A}$ is a constant.  It is worth noting that one of the three constants determining the value of the connection correlation tensor $Z^{\alpha}{}_{\beta\mu}{}^{\gamma}{}_{\delta\nu}$ is determined by the constraints used in determining the affine deformation tensor.  The other two constants, $h_2$ and $b_1$, are not determined.

\subsection{The Averaged Einstein Field Equations}

The effective energy density and pressure due to the connection correlation tensor are
\begin{eqnarray}
\kappa\rho_{grav}&=& 6(H_1+H_2+H_3+H_5)\\
\kappa p_{grav}&=& 6(H_1-H_2-H_3-H_5)\nonumber\\
&&\qquad -4(H_3-H_4+H_7)
\end{eqnarray}
Substituting the solution found in equations (\ref{sol1},\ref{sol2},\ref{sol3},\ref{sol4}) we find that the effective energy density and pressure due to the connection correlation tensor becomes
\begin{equation}
\kappa\rho_{grav} = -3\frac{{\mathcal A}^2}{R^2}, \qquad \kappa p_{grav} = \frac{{\mathcal A}^2}{R^2}.
\end{equation}
and the averaged Einstein Field equations, equation (\ref{MG_eqns}) become
\begin{eqnarray}
3\frac{(R')^2}{R^4}&=& \kappa \rho_{mat}+ \kappa \rho_{grav},\nonumber \\
\frac{(R')^2}{R^4}-2\frac{R''}{R^3}&=& \kappa p_{mat} + \kappa p_{grav}.\label{MGeqns}
\end{eqnarray}
We know that $\langle {\bf t}_{\beta
}^{\alpha {\rm (micro)}}\rangle$ is also conserved and yields the usual conservation equation
\begin{equation}
\rho_{mat}'=-3\frac{R'}{R}(\rho_{mat}+p_{mat})
\end{equation}

The sign of $\rho_{grav}$ is determined explicitly from the Macroscopic Gravity equations that determine the affine deformation tensor.  One also observes that the effective total energy density is less than the energy density of the averaged matter, that is, $\rho_{eff}=\rho_{mat} + \rho_{grav} < \rho_{mat}$.  We observe that the effect of the connection correlations within this model and with our assumptions is to essentially add a positive pressure term to the classical (non-averaged) Einstein Field equations.  Furthermore, the effective acceleration due to matter and connection correlations due to averaging remains the same as that of just the averaged matter, $\rho_{eff}+3p_{eff}=\rho_{mat}+\rho_{grav}+3p_{mat}+3p_{grav}=\rho_{mat}+3p_{mat}$. Effectively, the connection correlations in Macroscopic Gravity, adds a positive spatial curvature term to the classical field equations of General Relativity without changing the underlying geometry.


\section{Conclusions}

The initial presentation of this averaged cosmological solution without many details can be found in \cite{ColeyPelavasZalaletdinvov2005}.  Due to the nature of the letter, not all of the equations describing Macroscopic Gravity are resolved, nor sufficient details given for someone to repeat the calculations.  Here, we develop a formalism in which one is able to repeat the necessary calculations and have used this formalism to present a complete solution to Macroscopic Gravity equations detailing all components of both the connection correlation tensor $Z^{\alpha}{}_{\beta\mu}{}^{\gamma}{}_{\delta\nu}$ and the affine deformation tensor $A^{\alpha}{}_{\beta\gamma}$.  With some clearly stated assumptions, (connection correlation tensor $Z^{\alpha}{}_{\beta\mu}{}^{\gamma}{}_{\delta\nu}$ and the affine deformation tensor $A^{\alpha}{}_{\beta\gamma}$ are invariant under the same group of isometries as the macroscopic metric, and the electric part of the connection correlation tensor is zero), a self-consistent and detailed solution to all of the Macroscopic Gravity equations was constructed.  The connection correlation tensor was found to have three degrees of freedom, and more remarkably the sign of the effective energy density due to the connection correlations, $\rho_{grav}$, is determined by the Macroscopic Gravity equations determining the affine deformation tensor.  It is this particular analysis of the affine deformation tensor that is not found in \cite{ColeyPelavasZalaletdinvov2005}.   The energy density due to connection correlations appears as a negative correction to the total energy density, but has no effect on the effective acceleration of a co-moving observer, that is, $\rho_{grav}+3p_{grav}=0$.  Indeed, the effective energy density due to connection correlations appears as a positive spatial curvature correction in the averaged Einstein Field Equations.  This is in stark contrast to the conclusion of Futamase \cite{Futamase}, who, using a perturbative approach to the averaging problem determined that the correction to the averaged Einstein Field Equations is a negative spatial curvature term.

With this complete solution to the Macroscopic Gravity equations, we observe that the geometry is one aspect of the problem, but the dynamics of the geometry due to gravitational correlations is another \cite{ColeyPelavasZalaletdinvov2005,Zalaletdinov2008}.  This fundamental difference between geometry of the spacetime and the dynamics within will most certainly have consequences on cosmological observations.  Indeed, spatial curvature, whether a result of backreaction or not, can lead to a significant effect on the evolution of the universe and the measurement of cosmological parameters \cite{Curvature_issues}.  Furthermore, recent discussions in \cite{null} argue that additional investigations are also required to analyze and interpret cosmological observations in the context of an averaging paradigm.

Here and in \cite{vandenHoogen2008}, it is observed that the sign of $\rho_{grav}$ is a direct consequence of the assumption that ${\overline r}^{\alpha}{}_{\beta\mu\nu}$ yields another curvature tensor with non-metric connection $\Pi^{\alpha}{}_{\beta\gamma}$ for the macroscopic or averaged manifold.  It is not clear at the present time, the extent of the importance of this assumption within the entire scheme of Zalaletdinov's theory of Macroscopic Gravity.   For if this assumption is dropped, then the sign of $\rho_{grav}$ in the scenario investigated here can indeed be positive.

Another look at the differential cyclic constraints, equations (\ref{DE1}) and (\ref{DE2}), reveals that in the co-moving coordinate system used here, where the shear and vorticity of the four-velocity vanish, removing the invariance of $Z^{\alpha}{}_{\beta\mu}{}^{\gamma}{}_{\delta\nu}$ under a ${\mathcal G}_6$ will have no effect on the form of the solution, that is, each non-trivial component of $Z^\alpha{}_{\beta\mu}{}^\gamma{}_{\delta\nu}$ will remain a constant.   The effective energy density due to connection correlations will continue to appear as an $(\mbox{constant} \times R^{-2})$ correction in the averaged Einstein Field Equations of Macroscopic Gravity.  Therefore, in order to obtain non-spatial curvature like corrections, one would need to investigate the Macroscopic Gravity equations with a non-trivial electric component, and/or investigate different macroscopic spacetime geometries.  One other geometry has been investigated in \cite{vandenHoogen2008}, but $ZE^{\alpha}{}_{\beta}{}^{\gamma}{}_{\delta\nu}=0$ and once again the correction/backreaction term appears as a curvature term.

One may think that setting $ZH^{\alpha}{}_{\beta}{}^{\gamma}{}_{\delta\nu}=0$ and determining the effects of the purely electric part, $ZE^{\alpha}{}_{\beta}{}^{\gamma}{}_{\delta\nu}$, on these models may be a possible scenario worth pursuing.  However, calculations show that if $Z^{\alpha}{}_{\beta\mu}{}^{\gamma}{}_{\delta\nu}$ is invariant under the same group of isometries as the macroscopic metric and if $ZH^{\alpha}{}_{\beta}{}^{\gamma}{}_{\delta\nu}=0$ then $ZE^{\alpha}{}_{\beta}{}^{\gamma}{}_{\delta\nu}=0$, and therefore all the connection correlations are trivial.

Although, the final result contained within this paper did not provide a viable and alternative explanation describing the effects of Dark Matter and/or Dark Energy, it does indeed illuminate an alternative path that may eventually lead one to a possible explanation.

\begin{acknowledgements}
 This research was supported by St. Francis Xavier University's Council on Research, funding from the W.F. James Chair, and an NSERC Discovery Grant.  The author wishes to acknowledge the contributions of Dr. Roustam Zalaletdinov who guided the preliminary stages of this research during his stay at St. Francis Xavier University as the W.F. James Chair in Pure and Applied Science. The author would also like to thank Dr. Nicos Pelavas for his technical advice.
\end{acknowledgements}


\section*{References}
\begin{thebibliography}{10}

\bibitem{ColeyPelavasZalaletdinvov2005} A.A. Coley, N. Pelavas, and R.M. Zalaletdinov, {\it Phys. Rev. Lett.}, {\bf 95},  151102, (2005).

\bibitem{WMAP} D. N. Spergel et al., [WMAP Collaboration], {\it Astrophys. J. Suppl.},  {\bf 148}, 175 (2003);
C. L. Bennett et al., {\it Astrophys. J. Suppl.}, {\bf 583}, 1 (2003).

\bibitem{DE} S. Perlmutter et al., {\it Astrophysical Journal}, {\bf 517}, 565 (1999);  A. G. Reiss et al., {\it Astrophysical Journal}, {\bf 116}, 1009, (1998).

\bibitem{backreaction} E. W. Kolb, S. Matarrese, and A. Riotto, {\it New J. Phys.}, {\bf 8}, 322, (2006); E. W. Kolb, S. Matarrese, A. Notari and A. Riotto, {\it Phys. Rev. D}, {\bf 71}, 023524, (2005);  S. R\"as\"anen, {\it J. Cosmol. Astropart. Phys.}, {\bf JCAP02(2004)003}, (2004);  S. R\"as\"anen, {\it Class. Quant. Grav.}, {\bf 23}, 1823-1835, (2006); R. A. Vanderveld, E. E. Flanagan and I. Wasserman, {\it Phys. Rev. D}, {\bf 76}, 083504 (2007); C. H. Chuang, J. A. Gu and W. Y. Hwang, {\it Class. Quant. Grav.}, {\bf 25}, 175001 (2008); J. Behrend, I. A. Brown and G. Robbers, {\it J. Cosmol. Astropart. Phys.}, 0801:013, (2008); T. Buchert, {\it Gen. Rel. Grav.} {\bf 40}, 467, (2008); T. Buchert and M. Carfora, {\it Class. Quantum Grav.}, {\bf 25} 195001, (2008); D.L. Wiltshire, {\it New J. Phys.}, {\bf 9}, 377, (2007); D.L. Wiltshire, {\it Phys. Rev. Lett.}, {\bf 99}, 251101, (2007) ;

\bibitem{Ellis} G.F.R. Ellis, in {\it General Relativity and Gravitation}, ed. B. Bertotti, F. de Felice and A. Pascolini (Dordrecht: Reidel, 1984), 215;  G.F.R. Ellis, W. Stoeger, {\it Class. Quantum Grav.}, {\bf 4}, 1697, (1987).

\bibitem{Zalaletdinov} R. M. Zalaletdinov, {\it Gen. Rel. Grav.}, {\bf  24}, 1015-1031, (1992); R. M. Zalaletdinov, {\it Gen. Rel. Grav.}, {\bf 25}, 673, (1993);

\bibitem{Zalaletdinov2} R. M. Zalaletdinov, {\it Bull. Astron. Soc. India}, {\bf 25}, 401, (1997); R. M. Zalaletdinov, {\it Hadronic Journal}, {\bf 21}, 170, (1998); R. M. Zalaletdinov, ``Averaging Problem in Cosmology and Macroscopic Gravity'', preprint arXiv:gr-qc/0701116 (2007).

\bibitem{Zalaletdinov2008} R.M. Zalaletdinov, {\it Int. J. Mod. Phys. A}, {\bf 23}, 1173-1181, (2008).

\bibitem{Krasinski} A. Krasi\'nski, Inhomogeneous Cosmological Models, (Cambridge University Press, Cambridge, 1997)

\bibitem{OLD_AVERAGE} M. F. Shirokov, I. Z. Fisher, {\it Sov. Astron. J.}, {\bf 6}, 699 (1963). Reprinted in: {\it Gen. Rel. Grav.}, {\bf 30} 1411-1427, (1998); T. W. Noonan, {\it Gen. Rel. Grav.}, {\bf 16}, 1103, (1984); T. W. Noonan, {\it Gen. Rel. Grav.}, {\bf 17}, 535, (1985);  R. A. Isaacson, {\it Phys. Rev.}, {\bf 166}, 1263, (1968); R. A. Isaacson, {\it Phys. Rev.}, {\bf 166}, 1272, (1968); S. Bildhauer and T. Futamase, {\it Gen. Rel. Grav.}, {\bf 23}, 1251-1264, (1991); N. V. Zotov and W. Stoeger, {\it Class. Quantum Grav.}, {\bf 9}, 1023 (1992); M. Kasai, {\it Phys. Rev. D}, {\bf 47}, 3214, (1993); J. P. Boersma, {\it Phys. Rev. D}, {\bf 57} (1998), 798;  T. Buchert, {\it Gen. Rel. Grav.}, {\bf 32}, 105-125 (2000);  T. Buchert, {\it Gen. Rel. Grav.}, {\bf 33}, 1381-1405, (2001);  J. Behrend,``Metric Renormalization in General Relativity'', preprint arXiv:gr-qc/0812.2859 (2008); F. Debbasch, {\it Eur. Phys. J. B}, {\bf 37}, 257-269, (2004).

\bibitem{Futamase} T. Futamase, {\it Phys. Rev. Lett.}, {\bf 61}, 2175, (1988); T. Futamase, {\it Phys. Rev. D}, {\bf 53}, 681, (1996).

\bibitem{COVAR_SPAT}  A. Paranjape and T.P. Singh, {\it Phys. Rev. D}, {\bf 76}, 044006 (2007); A. Paranjape, {\it Int. J. Mod. Phys. D} {\bf 17}, 597-601 (2008);

\bibitem{Coley_Pelavas} A. A. Coley and N. Pelavas, {\it Phys. Rev. D}, {\bf 74}, 087301, (2006); A. A. Coley and N. Pelavas, {\it Phys. Rev. D}, {\bf 75}, 043506, (2007).

\bibitem{vandenHoogen2008} R. J. van den Hoogen, {\it Gen. Rel. Grav.}, {\bf 40}, 2213-2227 (2008).

\bibitem{Zalaletdinov3} R. M. Zalaletdinov and R. J. van den Hoogen, ``The Macroscopic Gravity Equations for $Z_{E}$ and $Z_{B}$ Variables for the $G_6$ SHI$_0$ MG Cosmology (Jan 5, 2008)'', unpublished, (2008).

\bibitem{Stephani2004} H. Stephani, Relativity, (Cambridge University Press, Cambridge, 2004).

\bibitem{EllisvanElst} G.F.R. Ellis and H. van Elst, ``Cosmological Models, Carg\'ese Lectures 1998'', in Theoretical and Observational Cosmology, Ed. M Lachi\`eze-Rey, (Dordrecht: Kluwer, 1999), 1; Preprint gr-qc/9812046;  G.F.R. Ellis, ``Relativistic Cosmology'', in Cargese Lectures in Physics, Vol. 6, Ed. E. Schatzman (Gordon and Breach, New York, 1973).

\bibitem{MaartensMaharaj86} R. Maartens and S. D. Maharaj, {\it Class. Quant. Grav.}, {\bf 3}, 1005, (1986).

\bibitem{GRTENSOR} GRTensorII is a software package that runs within MAPLE.  It is distributed freely at http://grtensor.org.

\bibitem{Curvature_issues} A. A. Coley, ``Averaging and Cosmological Observations'', preprint arXiv:0704.1734 (2007);  C. Clarkson, M. Cortês and B. Bassett, {\it J. Cosmol. Astropart. Phys.}, {\bf JCAP08(2007)011 }, (2007).

\bibitem{null} S. R\"as\"anen, {\it J. Cosmol. Astropart. Phys.}, {\bf JCAP02(2009)011}, (2009); A. A. Coley, ``Cosmological Observations: Averaging on the Null Cone'', preprint arXiv:0905.2442 (2009); A. A. Coley, ``Null geodesics and observational cosmology'', preprint arXiv:0812.4565 (2008).

\end{thebibliography}
\end{document}